\documentclass{article}

\usepackage[preprint]{neurips_2020}

\usepackage[utf8]{inputenc} 
\usepackage[T1]{fontenc}    
\usepackage{hyperref}       
\usepackage{url}            
\usepackage{booktabs}       
\usepackage{amsfonts}       
\usepackage{nicefrac}       
\usepackage{microtype}      
\usepackage{graphicx}
\usepackage{amsmath}
\usepackage{algorithm}
\usepackage{algorithmic}
\usepackage{mathptmx}
\usepackage{amsmath}
\usepackage{amsthm}
\usepackage{booktabs}
\usepackage{amssymb}
\usepackage{multirow}
\usepackage{subfigure}
\usepackage{color}
\definecolor{gychen}{rgb}{0,0,1}
\definecolor{dong}{rgb}{1,0,0}
\usepackage{hyperref}

\newtheorem{theorem}{Theorem}

\title{Qatten: A General Framework for Cooperative Multiagent Reinforcement Learning}

\author{%
  Yaodong Yang$^1$, Jianye Hao$^1$\thanks{Corresponding author. Email: jianye.hao@tju.edu.cn}, Ben Liao$^2$
  Kun Shao$^3$, Guangyong Chen$^4$, Wulong Liu$^3$, Hongyao Tang$^1$\\
  $^1$College of Intelligence and Computing, Tianjin University, Tianjin, China\\
  $^2$Tencent Quantum Lab\\
  $^3$Huawei Noah's Ark Lab, China\\
  $^4$Shenzhen Institutes of Advanced Technology\\
  \texttt{} \\
}

\begin{document}

\maketitle

\begin{abstract}
In many real-world tasks, multiple agents must learn to coordinate with each other given their private observations and limited communication ability. Deep multiagent reinforcement learning (Deep-MARL) algorithms have shown superior performance in such challenging settings. One representative class of work is multiagent value decomposition, which decomposes the global shared multiagent Q-value $Q_{tot}$ into individual Q-values $Q^{i}$ to guide individuals' behaviors, i.e. VDN imposing an additive formation and QMIX adopting a monotonic assumption using an implicit mixing method. However, most of the previous efforts impose certain assumptions between $Q_{tot}$ and $Q^{i}$ and lack theoretical groundings. Besides, they do not explicitly consider the agent-level impact of individuals to the whole system when transforming individual $Q^{i}$s into $Q_{tot}$. In this paper, we theoretically derive a general formula of $Q_{tot}$ in terms of $Q^{i}$, based on which we can naturally implement a multi-head attention formation to approximate $Q_{tot}$, resulting in not only a refined representation of $Q_{tot}$ with an agent-level attention mechanism, but also a tractable maximization algorithm of decentralized policies. Extensive experiments demonstrate that our method outperforms state-of-the-art MARL methods on the widely adopted StarCraft benchmark across different scenarios, and attention analysis is further conducted with valuable insights.
\end{abstract}





\section{Introduction}
\label{section:introduction}
The cooperative multiagent reinforcement learning problem has achieved lots of attention in the last decade, where a system of agents learns towards coordinated policies to optimize the accumulated global rewards \cite{busoniu08,gupta_cooperative_2017,palmer_lenient_2018}. Complex tasks such as the coordination of autonomous vehicles \cite{Cao2012An}, optimizing the productivity of a factory in distributed logistics \cite{Ying2005Multi} and energy distribution, are often modeled as the cooperative multi-agent learning problems and have great application prospects with enormous commercial value. One natural way to address cooperative MARL problems is the centralized approach, which views the multiagent system (MAS) as a whole and solves it as a single-agent learning task. In such settings, existing reinforcement learning (RL) techniques can be leveraged to learn joint optimal policies based on agents’ joint observations and common rewards \cite{tan_multi-agent_1993}. However, the centralized approach usually scales not well, since the joint action space of agents grows exponentially as the number of agents increases. Furthermore, the partial observation limitation and communication constraints also necessitate the learning of decentralized policies, which conditions only on the local action-observation history of each agent \cite{foerster_counterfactual_2018}.

Another choice is the decentralized approach that each agent learns its policy. Letting individual agents learn concurrently based on the global reward (aka. independent learners) \cite{tan_multi-agent_1993} is the simplest option. However, it is shown to be difficult in even simple two-agent, single-state stochastic coordination problems. One main reason is that the global reward signal brings the non-stationarity that agents cannot distinguish between the stochasticity of the environment and the exploitative behaviors of other co-learners \cite{lowe_multi-agent_2017}, and thus may mistakenly update their policies. To mitigate this issue, decentralized policies can be learned in the centralized training with decentralized execution (CTDE) paradigm. 
CTDE learns a fully centralized value function based on the joint action and state information and then uses it to guide the optimization of decentralized policies, an approach taken by counterfactual multi-agent (COMA) policy gradients \cite{foerster_counterfactual_2018}, as well as work by \cite{gupta_cooperative_2017}. However, the centralized critic of COMA suffers difficulty in evaluating global Q-values from the joint state-action space especially when there are more than a handful of agents and is hard to give an accurate multiagent baseline \cite{schroeder_de_witt_multi-agent_2019}.

Different from the above methods, Value Decomposition Network (VDN) \cite{sunehag_value-decomposition_2018} represents $Q_{tot}$ as a sum of individual Q-values that condition only on individual observations and actions. Decentralized policies arise simply from each agent, which selects actions greedily based its $Q^{i}$. However, VDN supposes a strict additive assumption between $Q_{tot}$ and $Q^{i}$, and ignores any extra state information available during training. Later QMIX \cite{rashid_qmix_2018} is proposed to overcome VDN's limitations. QMIX employs a network that estimates joint action-values as a non-linear combination of per-agent values that condition on local observations. Besides, QMIX enforces that $Q_{tot}$ is monotonic in $Q^{i}$, which allows computationally tractable maximization of the joint action-value in off-policy learning. But QMIX performs an implicit mixing of $Q^{i}$ while regarding the mixing process as a black-box. Besides, when mixing individual $Q^{i}$s to $Q_{tot}$, QMIX uses weights directly produced from global features instead of accurately modeling the individual impact to the whole system at a per-agent level. Recently, QTRAN \cite{son_qtran_2019} is proposed to guarantee optimal decentralization inheriting the additive assumption while avoiding representation limitations introduced by VDN and QMIX. However, the constraints of QTRAN on the optimization problem involved are computationally intractable. Authors have to relax these constraints by two penalties thus deviating QTRAN from exact solutions \cite{mahajan_maven_2019}.

In this paper, we theoretically derive a decomposing formulation of $Q_{tot}$ by $Q^{i}$ for cooperative agents. Following the theoretical results, we propose a practical multi-head attention based Q-value mixing network (Qatten) to approximate the global Q-value. Qatten employs the key-value memory operation to explicitly measure the importance of each individual to the global system for transforming individual $Q^{i}$s to $Q_{tot}$ within a multi-head attention structure. Experiments on the challenging MARL benchmark show that our method obtains the best performance compared with other popular MARL algorithms. The attention analysis further shows how Qatten measures agents with different weights for mixing $Q^{i}$s into $Q_{tot}$ according to each agent's individual properties.



The remainder of this paper is organized as follows. We first introduce the Markov games and attention mechanism in Section~\ref{section:background}. Then in Section~\ref{section:theory}, we explain the explicit formula for local behavior of cooperative MARL and derive the mathematical relation between $Q_{tot}$ and $Q^{i}$. Next, in Section~\ref{section:method}, we present the framework of Qatten in detail. Furthermore, we validate Qatten in the challenging StarCraft II platform and give the mixing weight analysis in Section~\ref{section:experiment}. Finally, conclusions and future work are provided in Section~\ref{section:conclusion}.

\section{Background}
\label{section:background}
\subsection{Markov Games}
Markov games is a multi-agent extension of Markov Decision Processes \cite{littman_markov_1994}. They are defined by a set of states, $S$, action sets for each of $N$ agents, $A^{1},...,A^{N}$, a state transition function, $T: S \times A^{1} \times ... \times A^{N} \rightarrow P(S)$, which defines the probability distribution over possible next states, given the current state and actions for each agent, and a reward function for each agent that also depends on the global state and actions of all agents, $R^{i}:S \times A^{1} \times ... \times A^{N} \rightarrow \mathbb{R}$. We specially consider the partial observed Markov games, in which each agent $i$ receives a local observation $o^{i}: Z(S, i) \rightarrow O^{i}$. Each agent learns a policy $\pi^{i}: O^{i} \rightarrow P(A^{i})$ which maps each agent's observation to a distribution over its action set. Each agent learns a policy that maximizes its expected discounted returns, $J^{i}(\pi^{i})=\mathbb{E}_{a^{1} \sim \pi^{1},...,a^{N} \sim \pi^{N},s \sim T}[\sum_{t=0}^{\infty}\gamma^{t}r^{i}_{t}(s_{t},a^{1}_{t},...,a^{N}_{t})]$, where $\gamma\in [0,1]$ is the discounted factor. If all agents receive the same rewards ($R^{1}=...=R^{N}=R$), Markov games becomes fully-cooperative: a best-interest action of one agent is also a best-interest action of others \cite{matignon_independent_2012}. We consider the fully-cooperative partially observed Markov games in this paper. When agents make decisions, the history record of agent $i$'s action-observation, $\tau^{i}$, is usually used to replace the agent $i$'s observation $o^{i}$ and an Recurrent Neural Network (RNN) is added in the agent's local policy or Q-value function.

For multiagent value decomposition methods such as VDN and QMIX, an important concept for such methods is decentralisability, also called Individual-Global-Max (IGM) \cite{son_qtran_2019}, which asserts that $\exists Q^{i}$, such that the following holds:
\begin{equation}
\label{eq:igm}
    \arg\max_{\vec{a}}Q^{*}(\vec{\tau}, \vec{a}) = (\arg\max_{a^{i}} Q^{1}(\tau^{1},a^{1})...\arg\max_{a^{N}} Q^{N}(\tau^{N},a^{N})),
\end{equation}
where $\vec{\tau}$ is the joint agent action-observation histories.

\subsection{Attention Mechanism}
In recent years, the attention mechanism \cite{vaswani_attention_2017} has been widely used in various research fields and more and more works rely on the idea of the attention to deal with challenges in MAS. For example, MAAC \cite{iqbal_actor-attention-critic_2019} uses the self-attention to learn the critic for each agent by selectively paying attention to information from other agents while TarMAC \cite{pmlr-v97-das19a} allows targeted continuous communication between agents via a sender-receiver soft attention mechanism and multiple rounds of collaborative reasoning. An attention function can be described as mapping a query and a set of key-value pairs to an output, where the query $V_{Q}$, keys $V_{K}^{j}$, values, and output are all vectors. The output is computed as a weighted sum of values, where each weight $w_{j}$ assigned to each value is computed by a compatibility function of the query with the corresponding key.

\begin{equation}
\label{attention}
    w_{j} = \frac{exp(f(V_{Q}, V_{K}^{j}))}{\sum_{k}exp(f(V_{Q}, V_{K}^{k}))},
\end{equation}

where $f(V_{Q}, V_{K}^{j}))$ is the user-defined function to measure the importance of the corresponding value and the scaled dot-product is a common one. In practice, a multi-head structure is usually employed to allow the model to focus on information from different representation sub-spaces.


\section{Theoretical Analysis of Global and Individual Q-values}
\label{section:theory}
In this section, we first theoretically derive the expansion formula of each individual Q-value function and then investigate the non-linear combination of multiple individual Q-values into the global one under the multiagent value decomposition framework, which naturally inspires the development of our Qatten model as shown in the next section. 

Following the general framework of MARL, the Q-value $Q_{tot}(s,\vec{a})$ is a function on the state vector $s$ and joint actions $\vec{a}=(a^1,...,a^N).$ By applying the Implicit Function Theorem, $Q_{tot}$ can also be viewed as a function in terms of individual Q-value $Q^i$:
\begin{equation}
\label{eq:problem}
\begin{split}
    Q_{tot} & = Q_{tot}(s, Q^{1}, Q^{2}, ..., Q^{n}), \text{ where } Q^{i}=Q^{i}(s, a^{i}) \approx Q^{i}(\tau^{i}, a^{i}).
\end{split}
\end{equation}
Without any loss of generality, we assume that there is no 'independent' agent unrelated to the whole group, since an independent agent $i$ should not be considered as a group member, or it can be simply treated as an independent agent to optimize its policy. This means mathematically that the variation of individual Q-value $Q^i$ will have an impact (negative or positive) on the global Q-value $Q_{tot}$, that is $\frac{\partial{Q_{tot}}}{\partial{Q^i}}$ cannot be {\it identically} zero (whereas admitting {\it some} zero points is allowed),
\begin{equation}
 \frac{\partial{Q_{tot}}}{\partial{Q^i}}\neq 0.  \nonumber 
\end{equation}
To simplify our analysis, we investigate the local behaviors of $Q_{tot}$ and $Q^i$ nearly a maximum point $\vec{a}$ in the 
action space 
assuming the state $s$ is fixed.
Since the gradient $\frac{\partial{Q_{tot}}}{\partial{a^i}}$ vanishes 
at the  optimum point $\vec a_o$, we have
\begin{equation}
    \frac{\partial{Q_{tot}}}{\partial{a^i}}=\frac{\partial{Q_{tot}}}{\partial{Q^i}}\frac{\partial{Q^i}}{\partial{a^i}}=0.
\end{equation}
We conclude that 
\begin{equation}
     \frac{\partial{Q^i}}{\partial{a^i}}(a_o)=0.
\end{equation}
Consequently, we have local expansion of $Q^i(a^i)$ as follows,
\begin{equation}
    Q^i(a^i) = \alpha_i+\beta_i(a^i-a_o^i)^2+o((a^i-a_o^i)^2).
    \label{Eq_IndQ}
\end{equation}
where $\alpha_i$ and $\beta_i$ are constants. Thereafter, we can theoretically derive that the non-linear dependence of the global Q-value $Q_{tot}$ on individual Q-value $Q^i$ (near a maximal point $\vec a_o$) as shown in the following Theorem~\ref{theorem-multihead},.

\begin{theorem}
\label{theorem-multihead}
Assume that the action space is continuous and there is no independent agent. Then
there exist constants $c(s),\lambda_i(s)$ (depending on state $s$), such that the local expansion of $Q_{tot}$ admits the following form
\begin{equation}
\label{formula-multihead}
Q_{tot}(s,\vec{a}) \approx c(s) +\sum_{i,h} \lambda_{i,h}(s) Q^{i}(s,a^i),
\end{equation}
where
$\lambda_{i,h}$ is a linear functional of all partial derivatives $\frac{\partial^{h}Q_{tot}}{\partial Q^{i_1}...\partial Q^{i_h}}$ of order $h$, and decays super-exponentially fast in $h.$
\end{theorem}

A rigorous proof and analysis of Theorem~\ref{theorem-multihead} can be found in Appendix~\ref{appendix-proofs}, we briefly cover the proof idea here to keep our presentation intact.  At first glance, this theorem may be questioned since $Q_{tot}$ as a functional of all the $Q^i$s absolutely contains non-linear terms, such as $Q^iQ^j$. To address this issue, we multiple $Q^iQ^j$ directly and find that we can reformulate $Q^iQ^j$ with first-order terms $Q^{i}$ and $Q^{j}$. In the multiplication result of $Q^iQ^j$, $\alpha_i\alpha_j$ is a constant term;
The term $\alpha_i\beta_j(a^j-a_o^j)^2=\alpha_i (Q^j-\alpha_j)$ and thus contributes a linear term $\alpha_iQ^j$ and a constant term $-\alpha_i\alpha_j$, and so is the term $\beta_i(a^i-a_o^i)^2\alpha_j$; 
Finally, the fourth-order term $\beta_i(a^i-a_o^i)^2\beta_j(a^j-a_o^j)^2$ is negligible.
In this simple computation, the major part of the second-order term $Q^iQ^j$ is the same as the major part of a linear term $Q^i$ and $Q^j$, other non-linear higher-order terms follow the similar approximating process. Thus, we could just use individual Q-values to effectively approach the global Q-value. 

Eq.~(\ref{formula-multihead}) appears to be linear, yet contains the non-linear information since the head coefficient $\lambda_{i,h}$ is a function of all partial derivatives of order $h$, and corresponds to all cross terms $Q^{i_1}...Q^{i_h}$ of order $h$ (e.g. $\lambda_{i,2}$ corresponds to second order non-linearity $Q^iQ^j$).
When we implement Theorem~\ref{theorem-multihead} to design our multiagent value decomposition network in realistic applications, we propose to adopt the attention mechanism as the universal function approximator \cite{yun2019transformers} to approximate the coefficients $\lambda_{i,h}(s)$ (upto a normalization factor). The Universal Approximation Theorem for Transformer proved in \cite{yun2019transformers} mathematically justifies our choice, where attention mechanism is shown to be the main force of the Transformer's approximation ability. Next, we introduce our practical framework in detail based on the decomposition formation and validate its effectiveness in the experiment section.

\section{The Practical Implementation of Qatten}
\label{section:method}

\begin{figure*}[htbp]
\centering
{\includegraphics[height=2.8in,angle=0]{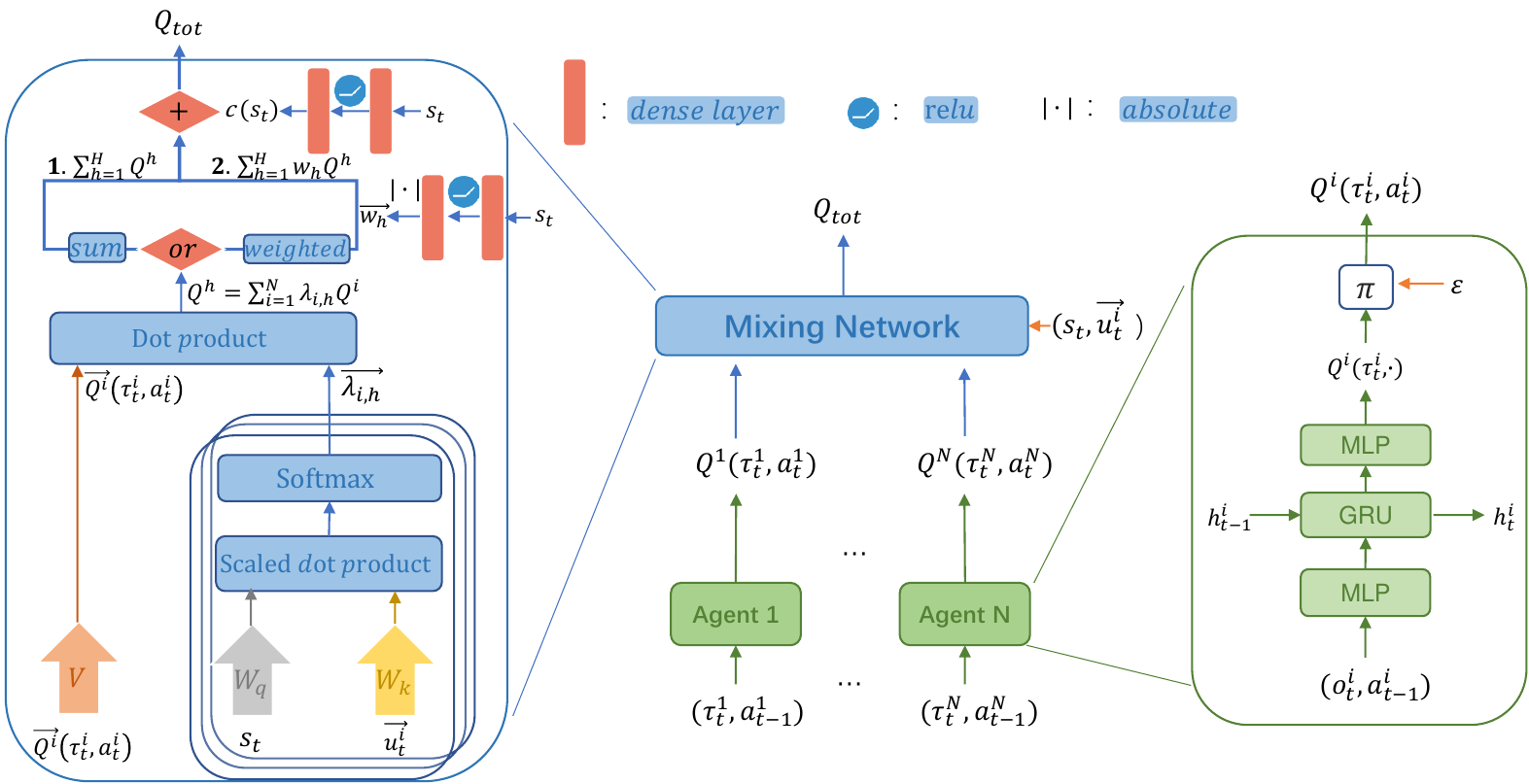}}
\caption{The overall architecture of Qatten. The right is agent $i$'s recurrent deep Q-network, which receives the action-observation history record $\tau^{i}$ (last hidden states $h_{t-1}^{i}$, current local observations $o_{t}^{i}$ and last action $a_{t-1}^{i}$). The left is the mixing network of Qatten, which mixes $\vec{Q^i}(\tau^{i}_{t}, a^{i}_{t})$ together with $s_t$ and $\vec{u^{i}_{t}}$. In general, $s$ is the global state and $u^{i}$ is the agent $i$'s individual features like its position.}
\label{figure:attqmix}
\end{figure*}

In this section, we propose the Q-value Attention network (Qatten), a practical deep Q-value decomposition network following the above decomposition formation in Eq.~(\ref{formula-multihead}). Figure~\ref{figure:attqmix} illustrates the overall architecture, which consists of agents' recurrent Q-value networks representing each agent's individual value function $Q^{i}(\tau^{i},a^{i})$ and the refined attention based value-mixing network to 
model the agent-level individual impacts while transforming individual Q-values to $Q_{tot}$. Besides $Q^{i}$s, the global information (including $s$ and $\vec{u^{i}}$) is also fed into the attention-based mixing network.

We start from the decomposition formation of $Q_{tot}$ and $Q^{i}$. Letting the outer sum over $h$ in Eq.(~\ref{formula-multihead}), we have 
\begin{equation}
\label{eq:qtot_qi}
\begin{split}
    & Q_{tot} \approx c(s) + \sum_{h=1}^{H} \sum_{i=1}^{N} \lambda_{i,h}(s) Q^{i}. \\
\end{split}
\end{equation}
For each $h$, the inner weighted sum operation can be implemented using the differentiable key-value memory model \cite{graves_neural_2014,oh_control_2016} to approximate the coefficients and establish the relations from the individuals to the global.
As mentioned in the previous section, differentiable key-value memory model admits powerful function approximation capability \cite{yun2019transformers}.
This is different from MAAC \cite{iqbal_actor-attention-critic_2019}, which uses the self-attention to learn the critic for each agent by selectively paying attention to information from other agents. Here we do not perform self-attention among each pair of agents, but use this mechanism to help the whole system to model each individual agent's impact at a per-agent level.
Specifically, we pass the similarity value between the global state's embedding vector $e_{s}(s)$ and the individual features' embedding vector $e_{i}(u^{i})$ into a softmax.
\begin{equation}
\label{eq:attention}
    \lambda_{i, h} \propto \exp(e_{i}^{T}W_{k, h}^{T}W_{q, h}e_{s}),
\end{equation}
where $W_{q, h}$ transforms $e_{s}$ into the global query and $W_{k, h}$ transforms $e_{i}$ into the individual key. The $e_{s}$ and $e_{i}$ could be obtained by a one or two-layer embedding transformation for $s$ and $u^{i}$. Next, for the outer sum over $h$, we use multiple attention heads to implement the approximations of different orders of partial derivatives. By summing up the head Q-values $Q^{h}$ from different heads, we get
\begin{equation}
\label{eq:qtot_qh}
    Q_{tot} \approx c(s) + \sum_{h=1}^{H} Q^{h} \text{, where } Q^{h} = \sum_{i=1}^{N} \lambda_{i,h} Q^{i}.
\end{equation}
$H$ is the number of attention heads. Lastly, the first term $c(s)$ in Eq.~(\ref{formula-multihead}) could be learned by a neural network with the global state $s$ as the input.

Qatten naturally holds the monotonicity and then achieves IGM property between $Q_{tot}$ and $Q^{i}$. 
\begin{equation}
\label{eq:mono}
    \frac{\partial Q_{tot}}{\partial Q^{i}} \ge 0, \forall i \in \{1,2,...,N\},
\end{equation}
Thus, Qatten allows tractable maximization of the joint action-value in off-policy learning and guarantees consistency between the centralized and decentralized policies. 
\paragraph{Weighted Head Q-value} 
In previous descriptions, we directly add the Q-value contribution from different heads. To relax the weight boundary limitation imposed by the self-attention and increase Qatten's representation ability, we could assign weights $w_{h}$ for the Q-values from different heads. To ensure monotonicity, we retrieve these head Q-value weights with an absolute activation function from a two-layer feed-work network $f^{NN}$, which adjusts head Q-values based on global states $s$.
\begin{equation}
\label{eq:qtot_weighted}
    Q_{tot} \approx c(s) + \sum_{h=1}^{H} w_{h} \sum_{i=1}^{N} \lambda_{i,h} Q^{i},
\end{equation}
where $w_{h}=|f^{NN}(s)|_{h}$. Compared with directly summing up head Q-values, the weighted head Q-values relaxes the upper bound and lower bound of $Q_{tot}$ imposed by attention. Note that Eq.~(\ref{eq:qtot_weighted}) still satisfies the formation of Eq.~(\ref{formula-multihead}) as
\begin{equation}
\nonumber
    Q_{tot} \approx c(s) + \sum_{h=1}^{H} w_{h} \sum_{i=1}^{N} \lambda_{i,h} Q^{i} = c(s) + \sum_{h,i} w_{h} \lambda_{i,h} Q^{i} 
    = c(s) + \sum_{i,h} \hat{\lambda}_{i,h} Q^{i},
    \text{ for each $i$ and $h$, } \hat{\lambda}_{i,h} = w_{h} \lambda_{i,h}.
\end{equation}



\section{Experimental Evaluation}
\label{section:experiment}

\subsection{Settings}
In this section, we evaluate Qatten in the StarCraft II decentralized micromanagement tasks and use StarCraft Multi-Agent Challenge (SMAC) environment \cite{samvelyan_starcraft_2019} as our testbed, which has become a common-used benchmark for evaluating state-of-the-art MARL approaches such as COMA \cite{foerster_counterfactual_2018}, QMIX \cite{rashid_qmix_2018} and QTRAN \cite{son_qtran_2019}. We train multiple agents to control allied units respectively to beat the enemy, while the enemy units are controlled by a built-in handcrafted AI. All the results are averaged over 5 runs with different seeds. Training and evaluation schedules such as the testing episode number and training hyper-parameters are kept the same as QMIX in SMAC. For the attention part, the embedding dim for the query (global state $s$) and key (agent individual features $u^{i}$) is 32, and the head number is set to 4. More details are provided in the Appendix.

\begin{table}[htbp]
\centering
\small
\label{tab:senarios}
\caption{Maps in hard and super hard scenarios.}
\begin{tabular}{ccccc}
\hline
Name & Ally Units & Enemy Units & Type & Challenge \\ \hline
\begin{tabular}[c]{@{}c@{}}5m\_vs\_6m\end{tabular} & 5 Marines & 6 Marines & \begin{tabular}[c]{@{}c@{}}Asymmetric, Homogeneous\end{tabular} & \begin{tabular}[c]{@{}c@{}}Focusing fire\end{tabular} \\ \hline
\begin{tabular}[c]{@{}c@{}}3s\_vs\_5z\end{tabular} & 3 Stalkers & 5 Zealots & \begin{tabular}[c]{@{}c@{}}Asymmetric, Heterogeneous\end{tabular} & Kite enemy \\ \hline
\begin{tabular}[c]{@{}c@{}}2c\_vs\_64zg\end{tabular} & 2 Colossi & 64 Zerglings & \begin{tabular}[c]{@{}c@{}}Asymmetric, Heterogeneous\end{tabular} & Large action space \\ \hline
\begin{tabular}[c]{@{}c@{}}bane\_vs\_bane\end{tabular} & \begin{tabular}[c]{@{}c@{}}4 Banelings,\\ 20 Zerglings\end{tabular}  & \begin{tabular}[c]{@{}c@{}}4 Banelings,\\ 20 Zerglings\end{tabular} & \begin{tabular}[c]{@{}c@{}}Symmetric, Heterogeneous\end{tabular} & Baneling blasts properly \\ \hline
\begin{tabular}[c]{@{}c@{}}3s5z\_vs\_3s6z\end{tabular} & \begin{tabular}[c]{@{}c@{}}3 Stalkers,\\ 5 Zealots\end{tabular} & \begin{tabular}[c]{@{}c@{}}3 Stalkers,\\ 6 Zealots\end{tabular} & \begin{tabular}[c]{@{}c@{}}Asymmetric, Heterogeneous\end{tabular} & Medivac absorbs fire \\ \hline
\begin{tabular}[c]{@{}c@{}}MMM2\end{tabular} & \begin{tabular}[c]{@{}c@{}}1 Medivac,\\ 2 Marauders,\\ 7 Marines\end{tabular} & \begin{tabular}[c]{@{}c@{}}1 Medivac,\\ 3 Marauders,\\ 8 Marines\end{tabular} & \begin{tabular}[c]{@{}c@{}}Asymmetric, Heterogeneous\end{tabular} & Circuitous tactics \\ \hline
\end{tabular}
\end{table}

According to the map characters and learning performance of algorithms, SMAC divides these maps into three levels: easy scenarios, hard scenarios, and super hard scenarios. All map scenarios are of different agent number or agent type. The easy scenarios include 2s\_vs\_1sc, 2s3z, 3s5z, 1c3s5z and 10m\_vs\_11m. Here we briefly introduce the hard scenarios and super hard scenarios in Table 1 (the first four maps are hard and the last two are super hard) and more map details are in the Appendix.


\subsection{Validation}
\subsubsection{Easy Scenarios}
First, we validate Qatten on the easy scenarios. Table~\ref{tab:easy-results} shows the median test win rate of different algorithms. It shows that Qatten achieves competitive performance with QMIX and other popular methods. Qatten could master these easy tasks and also works well in heterogeneous and asymmetric settings. COMA's poor performance may result from the sample inefficient on-policy learning and the unstable baseline because of its naive critic structure \cite{iqbal_actor-attention-critic_2019}. Except for 2s\_vs\_1sc and 2s3z, IDL's win percentage is quite low as directly using global rewards to update policies brings the non-stationary, which becomes severe when the number of agents increases. QTRAN also performs not well, as the practical relaxations could impede the exactness of its updating \cite{mahajan_maven_2019}.

\begin{table}[htbp]
\centering
  \caption{Median performance of the test win percentage in easy scenarios.}
  \label{tab:easy-results}
  \begin{tabular}{ccccccc}
    \toprule
    \multicolumn{1}{c}{Scenario} & 
    \multicolumn{1}{c}{\begin{tabular}[c]{@{}c@{}} Qatten \end{tabular}} & 
    \multicolumn{1}{c}{\begin{tabular}[c]{@{}c@{}} QMIX\end{tabular}} & 
    \multicolumn{1}{c}{\begin{tabular}[c]{@{}c@{}} COMA \end{tabular}} & 
    \multicolumn{1}{c}{\begin{tabular}[c]{@{}c@{}} VDN \end{tabular}} & 
    \multicolumn{1}{c}{\begin{tabular}[c]{@{}c@{}} IDL \end{tabular}} &
    \multicolumn{1}{c}{\begin{tabular}[c]{@{}c@{}} QTRAN \end{tabular}} \\
    \midrule
    2s\_vs\_1sc & \textbf{100} & \textbf{100} & 97 & \textbf{100} & \textbf{100} & \textbf{100} \\
    2s3z & \textbf{97} & \textbf{97} & 34 & \textbf{97} & 75 & 83 \\
    3s5z & \textbf{94} & \textbf{94} & 0 & 84 & 9 & 13 \\
    1c3s5z & \textbf{97} & 94 & 23 & 84 & 11 & 67 \\
    10m\_vs\_11m & \textbf{97} & 94 & 5 & 94 & 19 & 59 \\
  \bottomrule
\end{tabular}
\end{table}

\begin{figure}[htbp]
\centering
\subfigure[\scriptsize{Map 5m\_vs\_6m}]{
\label{5m-6m}
\includegraphics[width=0.32\textwidth]{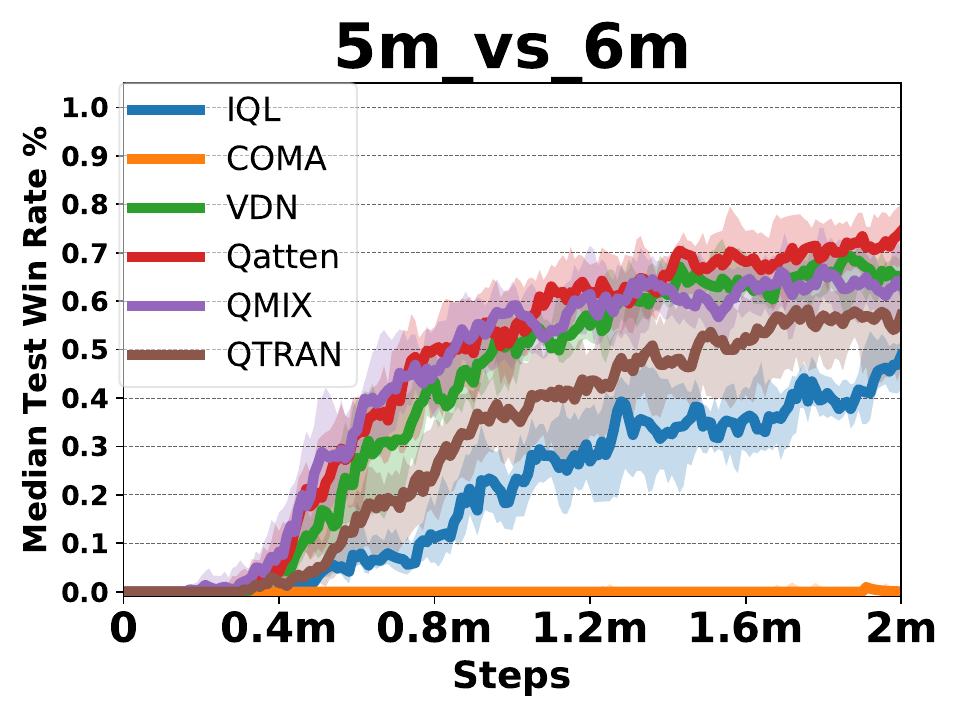}}
\subfigure[\scriptsize{Map 3s\_vs\_5z}]{
\label{3s-5z}
\includegraphics[width=0.32\textwidth]{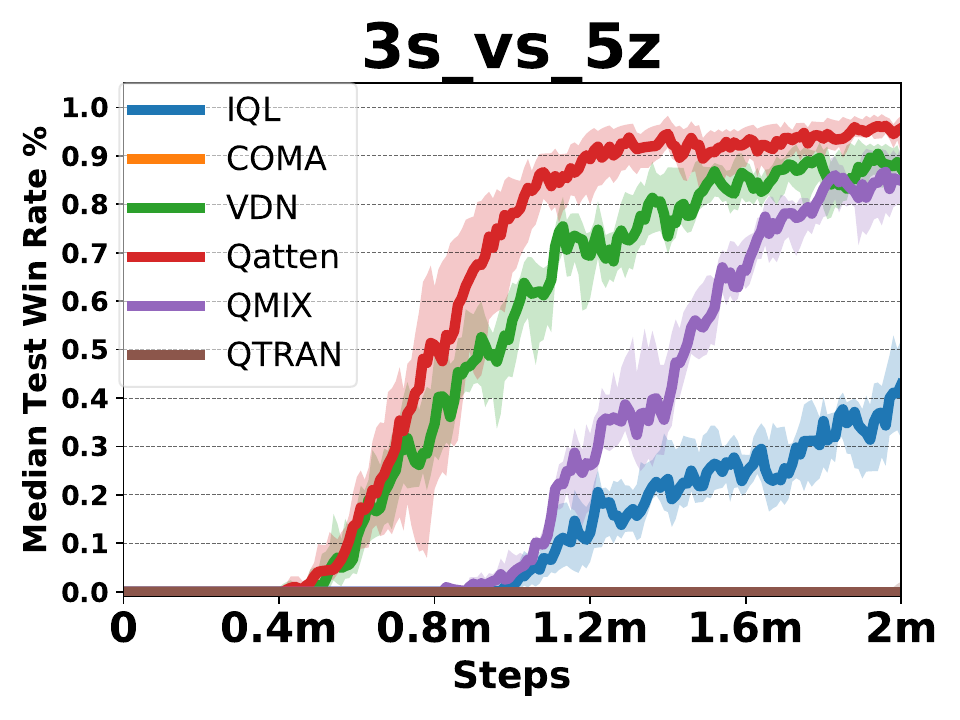}}
\subfigure[\scriptsize{Map bane\_vs\_bane}]{
\label{bane-bane}
\includegraphics[width=0.32\textwidth]{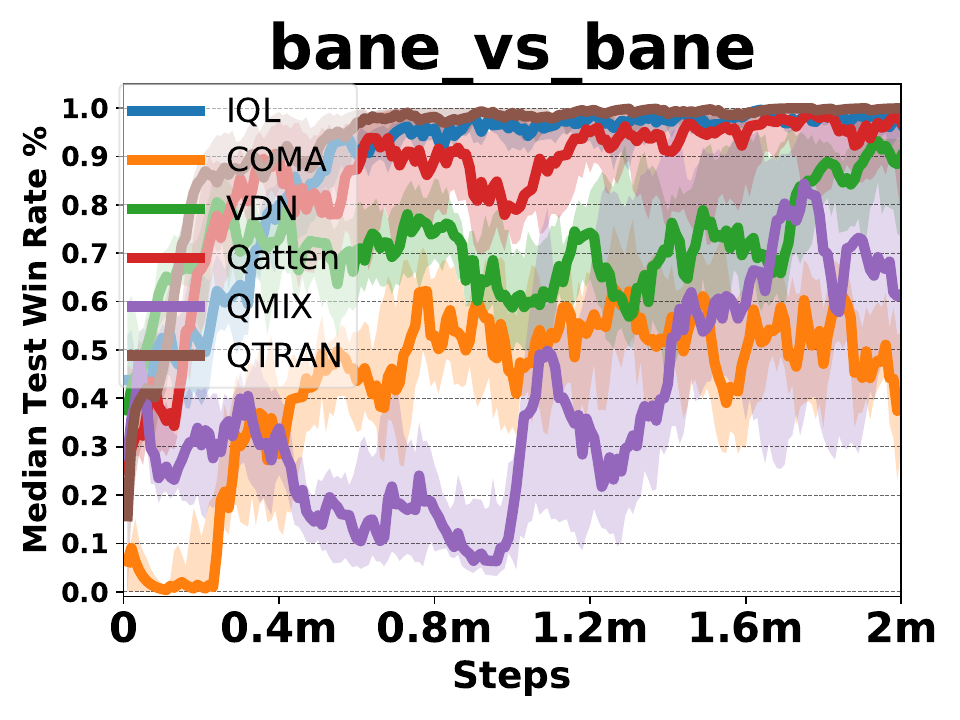}}
\subfigure[\scriptsize{Map 2c\_vs\_64zg}]{
\label{2c-64zg}
\includegraphics[width=0.32\textwidth]{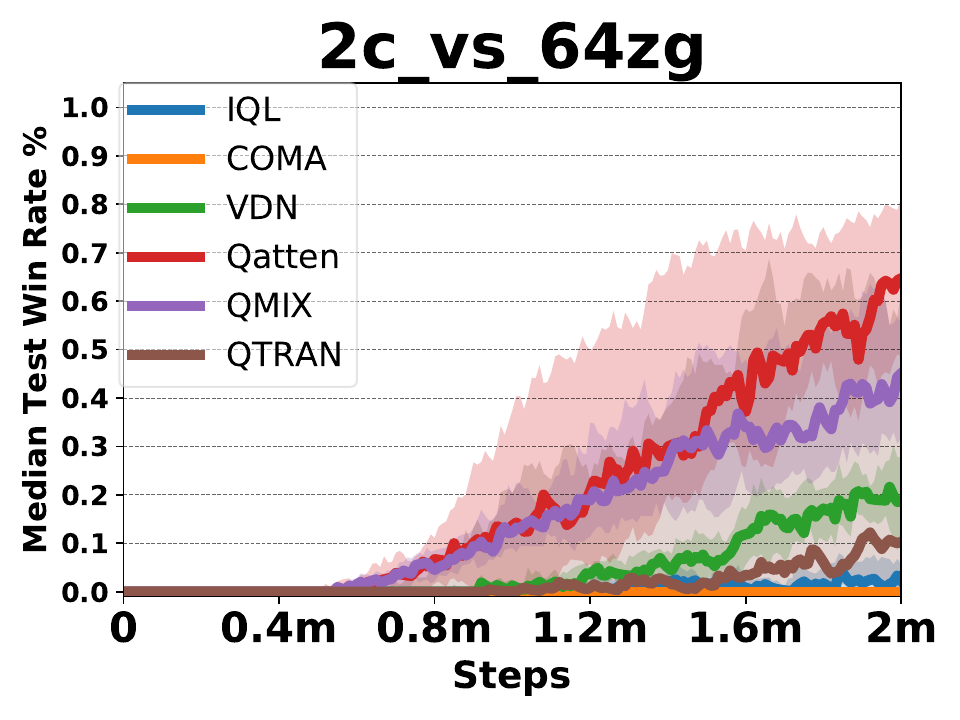}}
\subfigure[\scriptsize{Map MMM2}]{
\label{MMM2}
\includegraphics[width=0.32\textwidth]{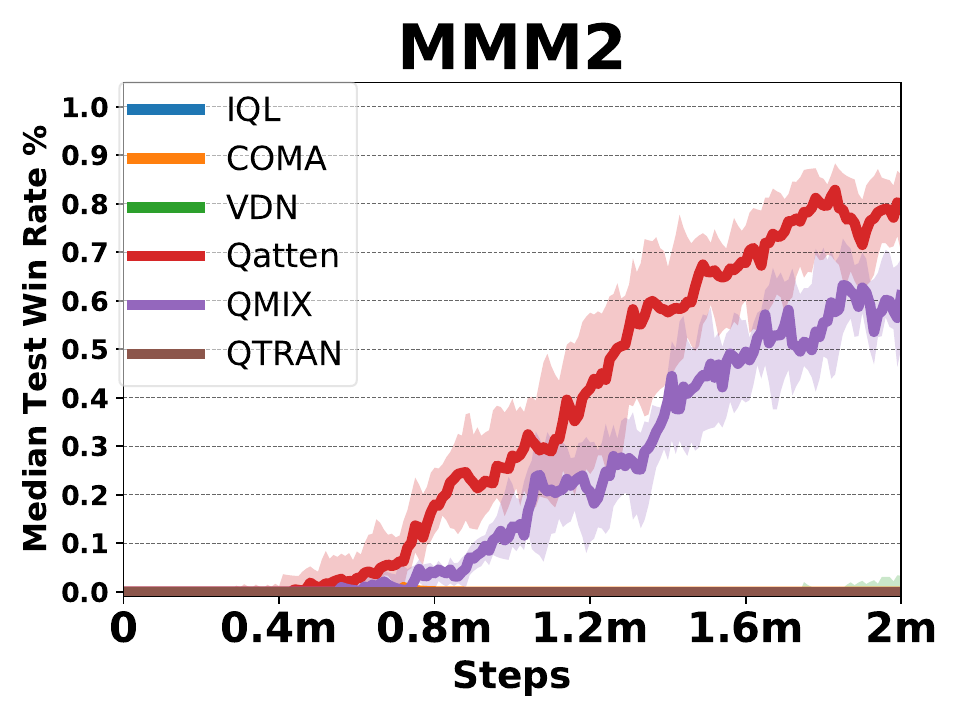}}
\subfigure[\scriptsize{Map 3s5z\_vs\_3s6z}]{
\label{3s5z_vs_3s6z}
\includegraphics[width=0.32\textwidth]{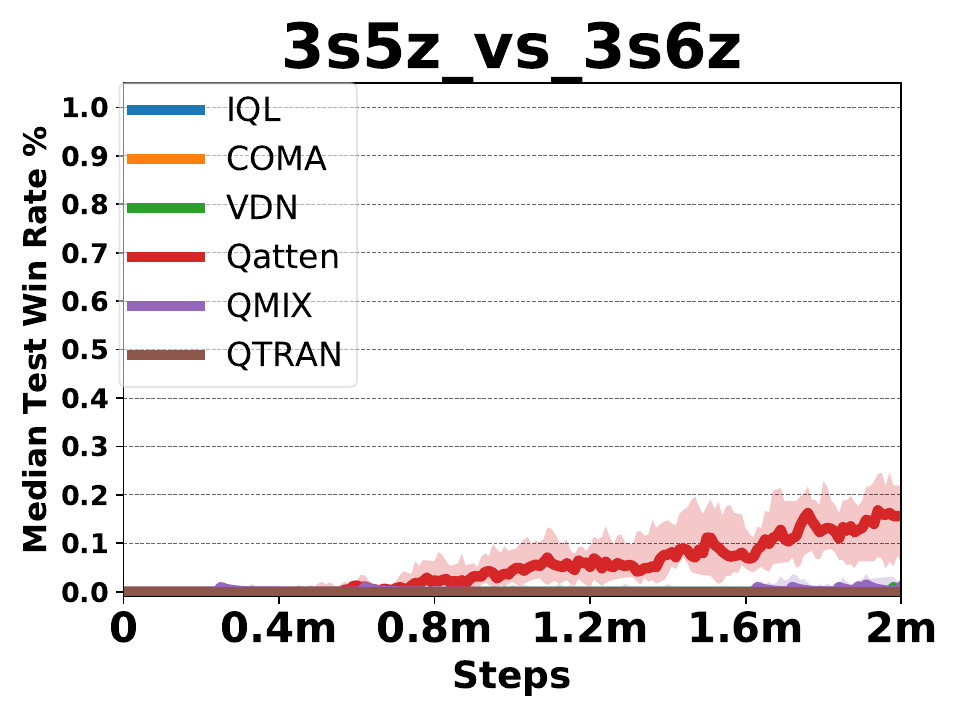}}
\caption{Median win percentage on the hard scenarios (a-d) and super hard scenarios (e-f).}
\label{figure:hard-results}
\end{figure}

\subsubsection{Hard Scenarios and Super Hard Scenarios}
Next, we test Qatten on the hard scenarios. Results are presented in Figure~\ref{5m-6m}-\ref{2c-64zg} as 25\%-75\% percentile is shaded. To summarize, these scenarios reflect different challenges, and existing approaches fail to achieve consistent performance while Qatten could consistently gain almost the best performance. For example, in bane\_vs\_bane which consists of 24 agents, Qatten performs much better than QMIX which cannot learn steadily. The reason is that the global Q-value heavily depends on the 4 Banelings among 24 agents as they are vital to winning the battle, thus Qatten only needs to concentrate on 4 Banelings while QMIX cannot easily find the appropriate formation for $Q_{tot}$ and $Q_{i}$. In the four maps, only Qatten consistently beats all other approaches, which validates the effectiveness of its attention based decomposition structure of $Q_{tot}$ and $Q^{i}$.

Finally, we test Qatten on the super hard scenarios as shown in Figure~\ref{MMM2}-\ref{3s5z_vs_3s6z} where 25\%-75\% percentile is shaded. Due to the difficulty and complexity, we augment Qatten with weighted head Q-values. In MMM2, Qatten masters the winning strategy by heading Medivac to absorb damage and then retreating it and exceeds QMIX by a large margin. The 3s5z\_vs\_3s6z is another more challenging scenario and all existing approaches fail. In contrast, our approach Qatten can win approximately 16\% after 2 million steps of training. 



\subsection{Ablation Study}
\begin{figure}[htbp]
\centering
\subfigure[\scriptsize{Map 2c\_vs\_64zg}]{
\label{2c-64zg-ablation}
\includegraphics[width=0.32\textwidth]{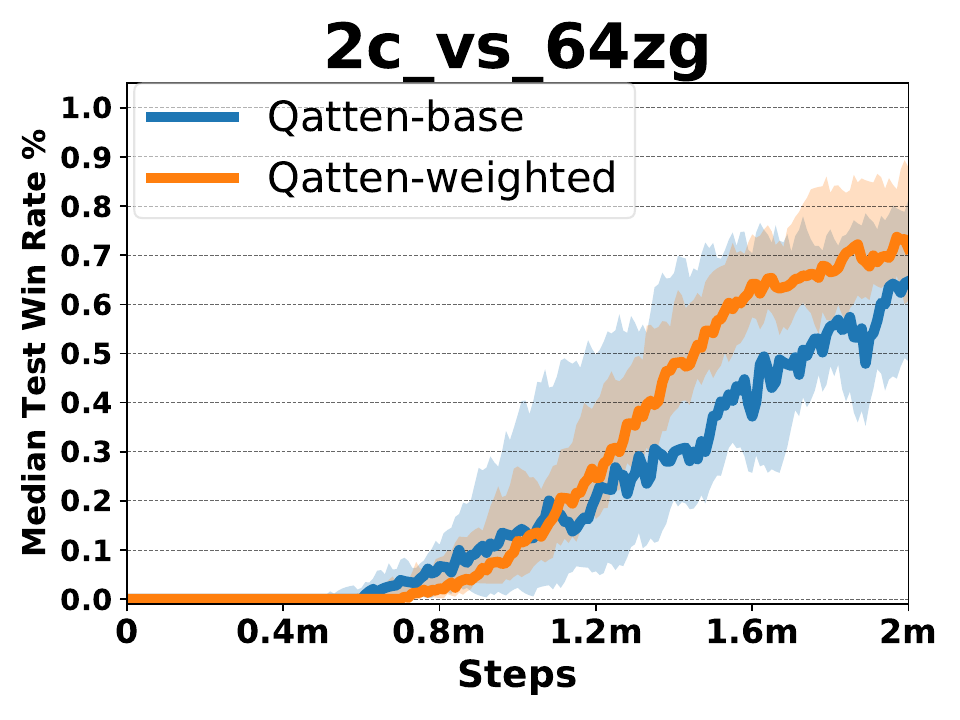}}
\subfigure[\scriptsize{Map MMM2}]{
\label{MMM2-ablation}
\includegraphics[width=0.32\textwidth]{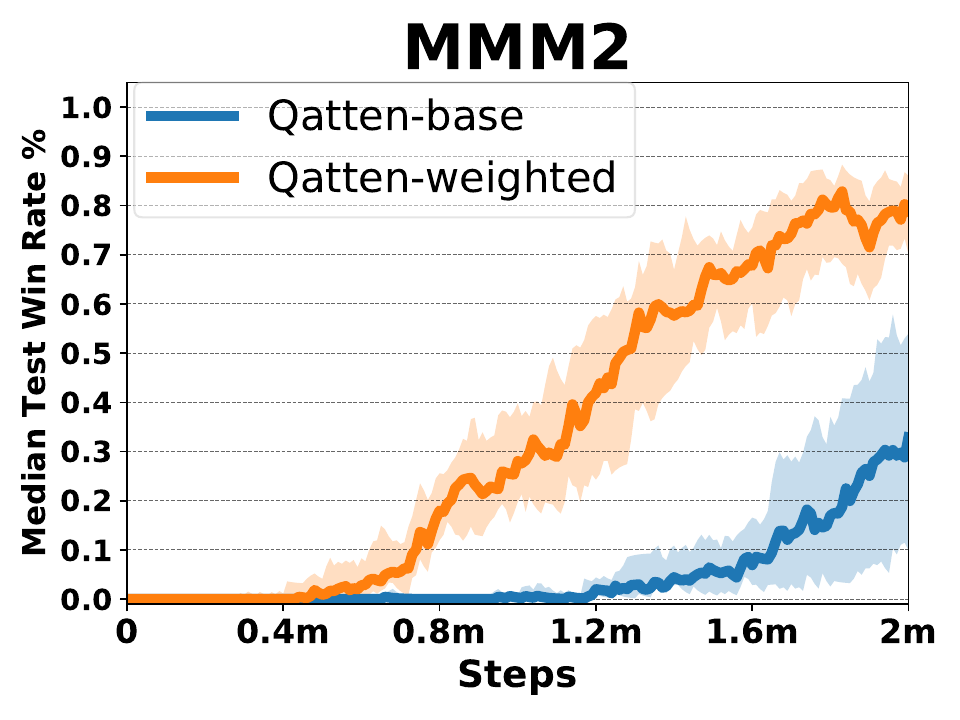}}
\subfigure[\scriptsize{Map 3s5z\_vs\_3s6z}]{
\label{3s5z-3s6z-ablation}
\includegraphics[width=0.32\textwidth]{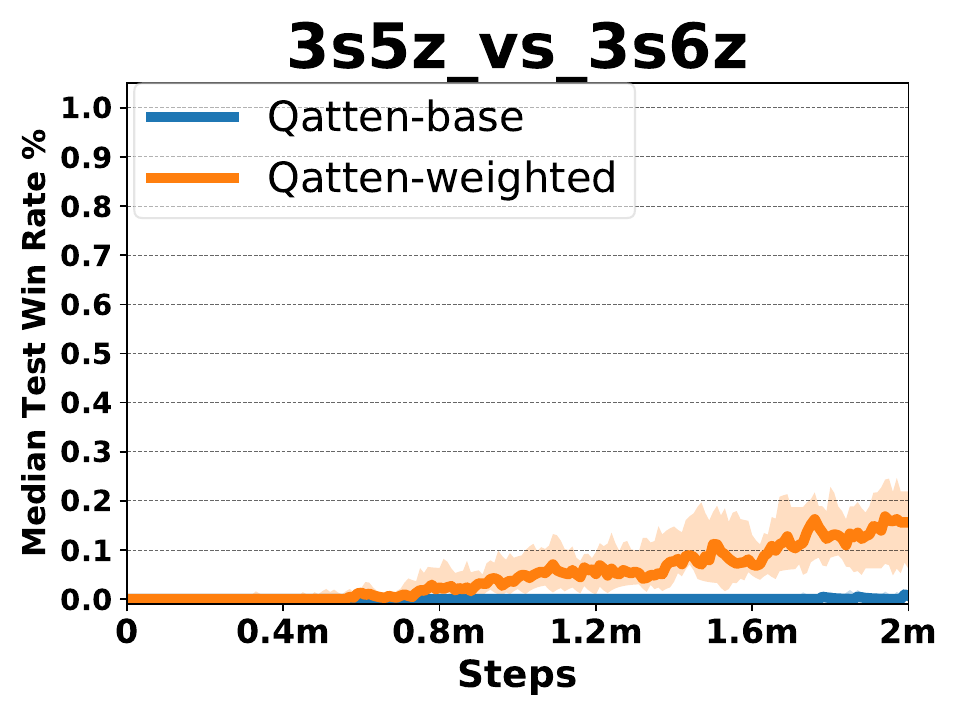}}
\caption{Ablation study of Qatten on three difficult scenarios.}
\label{figure:ablation}
\end{figure}

We investigate the influence of weighted head Q-values using three difficult scenarios as demonstrated in previous experiments. For clarity, the basic Qatten without weighted head Q-value is called Qatten-base while Qatten with weighted head Q-value is called Qatten-weighted. Figure~\ref{figure:ablation} shows the ablation results and 25\%-75\% percentile is shaded. As we can see, the weighted head Q-values can leverage the performance of Qatten-base on these difficult scenarios. It shows that this mechanism may capture sophisticated relations between $Q_{tot}$ and $Q^{i}$ with $w_{h}$ adjusting head weights flexibly to mitigate the boundedness imposed by the attention. In addition, in MMM2, $u^{i}$ is replaced by $(u^{i}, Q^{i})$. By the attention's softmax activation, 
it could enhance the non-linearity of $Q_{tot}$ and $Q^{i}$.

\subsection{Attention Analysis}
\begin{figure}[htbp]
\centering
\subfigure[\scriptsize{5m\_vs\_6m Attention Heat Map}]{
\label{5m_6m_heat}
\includegraphics[width=0.323\textwidth]{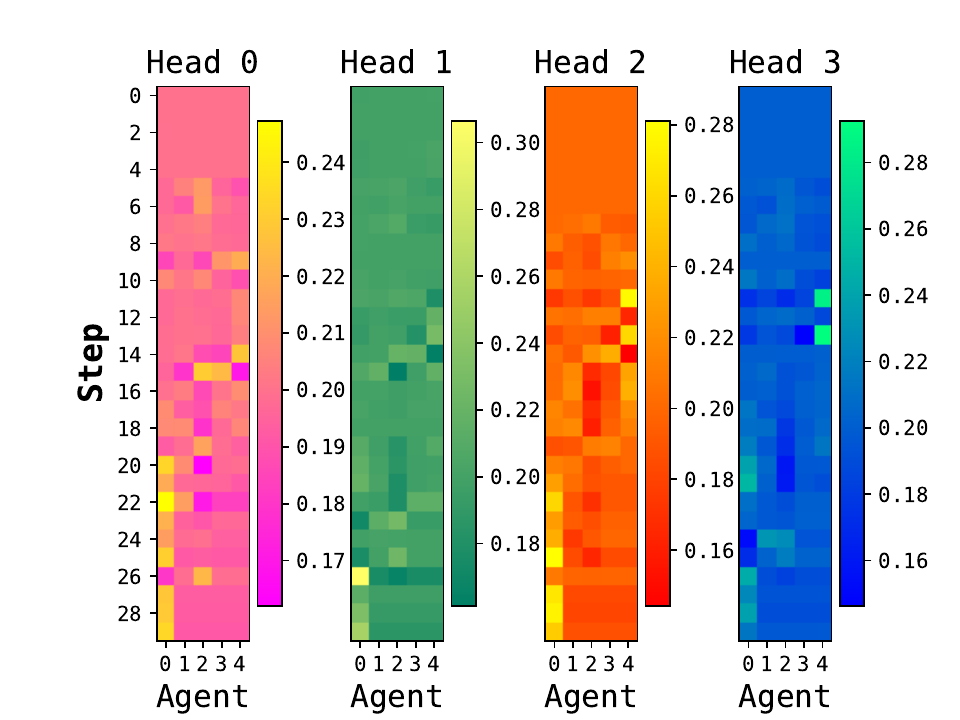}}
\subfigure[\scriptsize{5m\_vs\_6m Agent Health}]{
\label{5m_6m_health}
\includegraphics[width=0.323\textwidth]{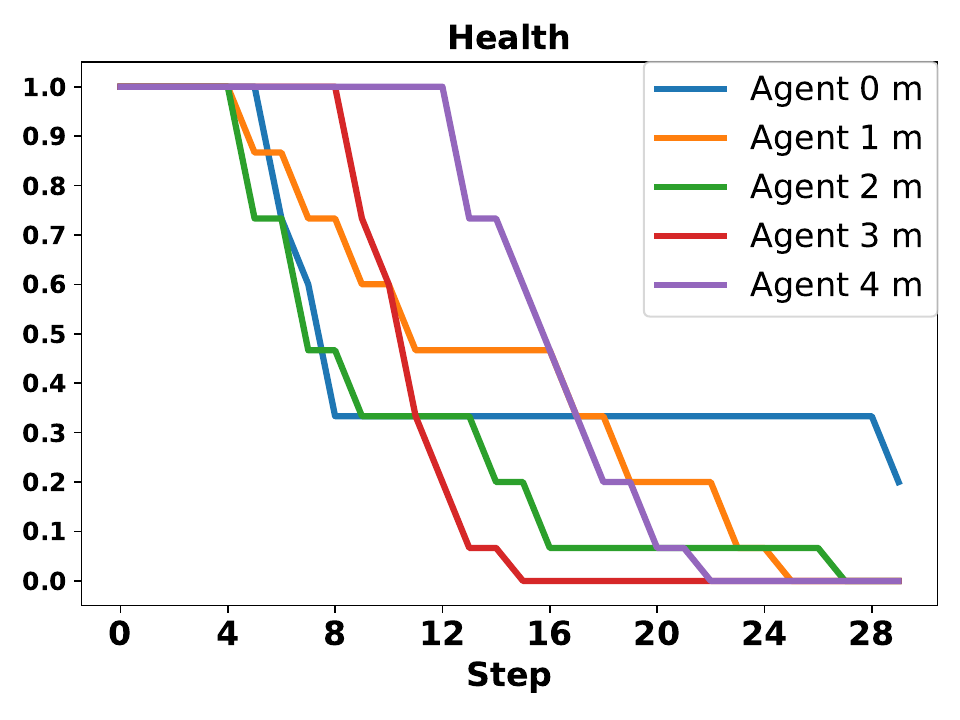}}
\subfigure[\scriptsize{3s5z\_vs\_3s6z Attention Heat Map}]{
\label{3s5z_3s6z_heat}
\includegraphics[width=0.323\textwidth]{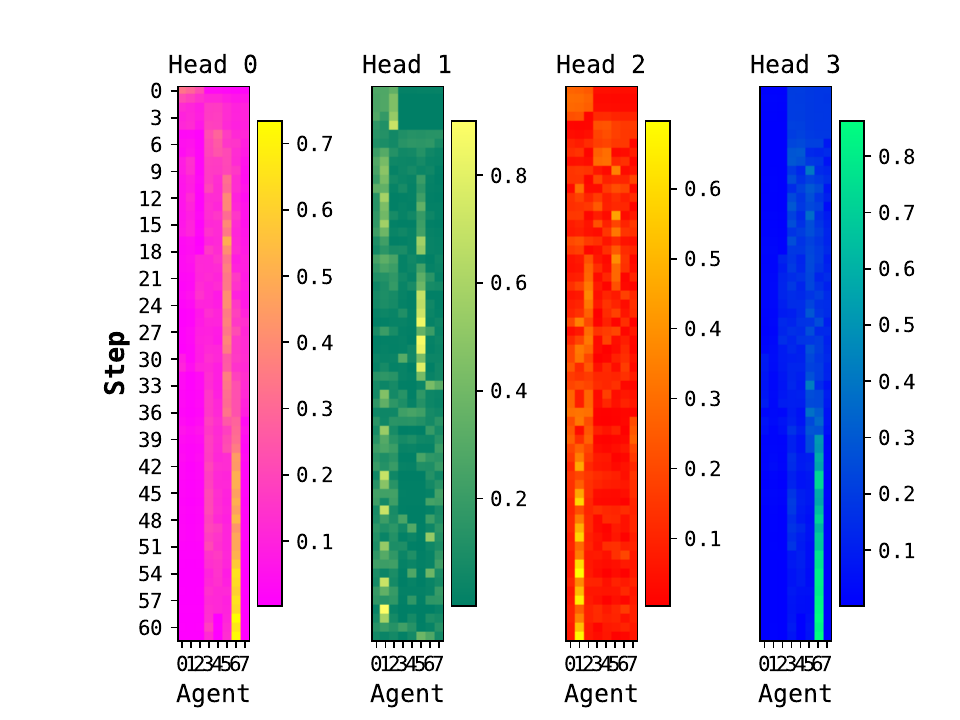}}
\label{figure:attention}
\caption{Attention weights on 5m\_vs\_6m and 3s5z\_vs\_3s6z. Steps increase from top to the bottom in the attention heat maps. Horizontal ordination indicates the agent id under each head.}
\end{figure}

We further visualize the attention weights ($\lambda_{i,h}$) on each step during a battle. We choose two representative maps 5m\_vs\_6m and 3s5z\_vs\_3s6z for illustration and the attention weight heat map of each head is shown in Figure~\ref{figure:attention}(a) and Figure~\ref{figure:attention}(c) respectively. For 5m\_vs\_6m, the most important trick to win is to avoid being killed and focusing fire to kill the enemy. From Figure~\ref{figure:attention}(a), we see allies have similar attention weights, which means that Qatten indicates an almost equally divided $Q_{tot}$ for agents. This is because each ally marine plays similar roles in this homogeneous scenario. This may also explain the fact that VDN could slightly outperform QMIX when the weights of each $Q^{i}$ is roughly equal. We also find some tendencies of Qatten. One main tendency is that the only alive unit (agent 0) receives the highest weight in the later steps of the episode. With the attention mechanism adjusting weights of $Q^{i}$ at a minor difference, Qatten performs better than VDN which assumes the weights of agents' individual Q-values are equal and static during the entire battle.

The weight difference is more apparent on the super hard map 3s5z\_vs\_3s6z as shown in Figure~\ref{figure:attention}(c). VDN and QMIX both fail at this challenging scenario while Qatten has the ability to approximate the sophisticated relations between $Q^{i}$ and $Q_{tot}$ at the agent level to better perceive each agent's role. For example, at the beginning of the battle, one kind of agents receive more attention than another kind under certain attention heads according to their type features as there exists two kinds of agents. A more detailed analysis based on the game background is in the Appendix~\ref{sec:extended_analysis}.

\section{Conclusion and Future Work}
\label{section:conclusion}
In this paper, we propose the novel Q-value Attention network for the multiagent Q-value decomposition problem. We first perform theoretical analysis of global and individual Q-values and derive their general decomposition formula. Following the decomposition formation, we introduce multi-head attention to establish the mixing network by explicit modeling the individual impact to the whole system at the agent level when transforming individual Q-values to $Q_{tot}$. Experiments on the challenging StarCraft MARL benchmark show that our method obtains the best performance on almost all maps and the attention analysis gives the intuitive explanations about the attention weights.

For future works, improving Qatten by combining with explicit exploration mechanisms on difficult MARL tasks is a straightforward direction. Besides, incorporating recent progresses of attention to adapt Qatten into large-scale settings where hundreds of agents exist is also promising.

\bibliographystyle{plain}
\bibliography{qatten}

\clearpage
\appendix

\section{Proofs}
\label{appendix-proofs}



{\bf Proof of  Theorem~\ref{theorem-multihead}:}
\begin{proof}

We expand $Q_{tot}$ in terms of $Q^i$
\begin{equation}
\label{eq-q-expand}
\begin{split}
Q_{tot}=  constant +  \sum_i \mu_i  Q^i + \sum_{ij} \mu_{ij} Q^i Q^j + ... + \sum_{i_1...i_k}\mu_{i_1...i_k}Q^{i_1}...Q^{i_k} + ...
\end{split}
\end{equation}
where
$$\mu_{i}=\frac{\partial Q_{tot}}{\partial Q^i},\mu_{ij}=\frac{1}{2}\frac{\partial^2 Q_{tot}}{\partial Q^i\partial Q^j},$$
and in general
$$\mu_{i_1...i_k}=\frac{1}{k!}\frac{\partial^k Q_{tot}}{\partial Q^{i_1}...\partial Q^{i_k}}.$$

Recall that for each individual Q value, we have local expansion
$$ Q^i(a^i)=\alpha_i+\beta_i(a^i-a_o^i)^2+o((a^i-a_o^i)^2).$$

Now we apply the equation above to the second order term in Equation~\ref{eq-q-expand}:
$$\sum_{ij}\mu_{ij} Q^i Q^j$$
$$=\sum_{ij}\mu_{ij} (\alpha_i+\beta_i(a^i-a_o^i)^2) (\alpha_j+\beta_j(a^j-a_o^j)^2)+o(\|a-a_o\|^2)$$
$$=\sum_{ij}\mu_{ij}\alpha_i\alpha_j + 2\sum_{ij}\mu_{ij}\alpha_j\beta_i(a^i-a^i_o)^2+o(\|a-a_o\|)^2$$
$$=\sum_{ij}\mu_{ij}\alpha_i\alpha_j + 2\sum_{ij}\mu_{ij}\alpha_j(Q^i-\alpha_i)+o(\|a-a_o\|)^2$$
$$=-\sum_{ij}\mu_{ij}\alpha_i\alpha_j + 2\sum_{ij}\mu_{ij}\alpha_jQ^i+o(\|a-a_o\|)^2$$
Therefore, we will take
$$\lambda_{i,2}=2\sum_{j}\mu_{ij}\alpha_j.$$

In general, we have
$$\mu_{i_1...i_k}=\frac{1}{k!}\frac{\partial^k Q_{tot}}{\partial Q^{i_1}...\partial Q^{i_k}}$$
and
$$\sum_{i_1,...,i_k}\mu_{i_1...i_k}Q^{i_1}...Q^{i_k}=-(k-1)\sum_{i_1,...,i_k}\mu_{i_1...i_k} + k\sum_{i_1,...,i_k}\mu_{i_1...i_k}\alpha_{i_1}...\alpha_{i_{k-1}}Q^{i_k}+o(\|a-a_o\|^2)$$
so we take
$$\lambda_{i,k}=k\sum_{i_1,...,i_k}\mu_{i_1...i_{k-1}i}\alpha_{i_1}...\alpha_{i_{k-1}}.$$

The convergence of the series $\sum_k \lambda_{i,k}$ only requires mild conditions, e.g. boundedness or even small growth of partial derivatives $\frac{\partial^k Q_{tot}}{\partial Q^{i_1}...\partial Q^{i_k}}$ in terms of $k.$
\end{proof}

We remark when we employ attention mechanism to approximate $\lambda_i$, $\lambda_{i,h}$ is modelled as single-head attention. 
Since $\lambda_{i,h}$ decays fast in $h$, we stop the series at $H$ (number of heads) for feasible computations in the practical implementation.

\section{Experimental Settings}
We follow the settings of SMAC \cite{samvelyan_starcraft_2019}, which could be referred in the SMAC paper. For clarity and completeness, we present these environment details again.

\subsection{States and Observations}
At each time step, agents receive local observations within their field of view. This encompasses information about the map within a circular area around each unit with a radius equal to the sight range, which is set to 9. The sight range makes the environment partially observable for agents. An agent can only observe others if they are both alive and located within its sight range. Hence, there is no way for agents to distinguish whether their teammates are far away or dead. If one unit (both for allies and enemies) is dead or unseen from another agent's observation, then its unit feature vector is reset to all zeros. The feature vector observed by each agent contains the following attributes for both allied and enemy units within the sight range: distance, relative x, relative y, health, shield, and unit type. If agents are homogeneous, the unit type feature will be omitted. All Protos units have shields, which serve as a source of protection to offset damage and can regenerate if no new damage is received. Lastly, agents can observe the terrain features surrounding them, in particular, the values of eight points at a fixed radius indicating height and walkability.

The global state is composed of the joint unit features of both ally and enemy soldiers. Specifically, the state vector includes the coordinates of all agents relative to the centre of the map, together with unit features present in the observations. Additionally, the state stores the energy/cooldown of the allied units based the unit property, which represents the minimum delay between attacks/healing. All features, both in the global state and in individual observations of agents, are normalized by their maximum values. In Qatten, the agent $i$'s individual features $u^{i}$ include its own position (relative coordination to the central point), health, cooldown, the possible shield (if available) and the possible type feature (in heterogeneous scenarios), which is the part of the global state.

\subsection{Action Space}
The discrete set of actions which agents are allowed to take consists of move[direction], attack[enemy id], stop and no-op. Dead agents can only take no-op action while live agents cannot. Agents can only move with a fixed movement amount 2 in four directions: north, south, east, or west. To ensure decentralization of the task, agents are restricted to use the attack[enemy id] action only towards enemies in their shooting range. This additionally constrains the ability of the units to use the built-in attack-move macro-actions on the enemies that are far away. The shooting range is set to be 6 for all agents. Having a larger sight range than a shooting range allows agents to make use of the move commands before starting to fire. The unit behavior of automatically responding to enemy fire without being explicitly ordered is also disabled. As healer units, Medivacs use heal[agent\_id] actions instead of attack[enemy\_id].

\subsection{Rewards}
At each time step, the agents receive a joint reward equal to the total damage dealt on the enemy units. In addition, agents receive a bonus of 10 points after killing each opponent, and 200 points after killing all opponents for winning the battle. The rewards are scaled so that the maximum cumulative reward achievable in each scenario is around 20.

\subsection{Training Configurations}
The training time is ranging from about 8 hours to 18 hours on these maps (GPU Nvidia RTX 2080 and CPU AMD Ryzen Threadripper 2920X), which is based on the agent numbers and map features of each map. The number of the total training steps is about 2 million and every 10 thousand steps we train and test the model. When training, a batch of 32 episodes are retrieved from the replay buffer which contains the most recent 5000 episodes. We use $\epsilon$-greedy policy for exploration. The starting exploration rate is set to 1 and the end exploration rate is 0.05. Exploration rate decays linearly at the first 50 thousand steps. We keep the default configurations of environment parameters.

\begin{table}[htbp]
\footnotesize
\centering
\caption{The network configurations of Qatten's mixing network.}
\label{tab:para-qatten}
\begin{tabular}{c|c}
\hline
\begin{tabular}[c]{@{}c@{}}Qatten mixing\\ network configurations\end{tabular}           & Value \\ \hline
Query embedding layer number     
& 2     \\ \hline
\begin{tabular}[c]{@{}c@{}}Unit number in\\ query embedding layer 1\end{tabular}         & 64    \\ \hline
\begin{tabular}[c]{@{}c@{}}Activation after\\ query embedding layer 1\end{tabular}          & Relu  \\ \hline
\begin{tabular}[c]{@{}c@{}}Unit number in\\ query embedding layer 2\end{tabular}         & 32    \\ \hline
Key embedding layer number                                                        & 1     \\ \hline
\begin{tabular}[c]{@{}c@{}}Unit number in\\ key embedding layer 1\end{tabular}           & 32    \\ \hline
Head weight layer number                                                          & 2     \\ \hline
\begin{tabular}[c]{@{}c@{}}Unit number in\\ head weight layer 1\end{tabular}             & 64    \\ \hline
\begin{tabular}[c]{@{}c@{}}Activation after\\ head weight layer 1\end{tabular}    & Relu  \\ \hline
\begin{tabular}[c]{@{}c@{}}Unit number in\\ head weight layer 2\end{tabular}             & 4     \\ \hline
Attention head number                                                             & 4     \\ \hline
Constant value layer number                                                       & 2     \\ \hline
\begin{tabular}[c]{@{}c@{}}Unit number in \\ constant value layer 1\end{tabular}         & 32    \\ \hline
\begin{tabular}[c]{@{}c@{}}Activation after\\ constant value layer 1\end{tabular} & Relu  \\ \hline
\begin{tabular}[c]{@{}c@{}}Unit number in \\ constant value layer 2\end{tabular}         & 1     \\ \hline
\end{tabular}
\end{table}

\subsection{Mixing Network Hyper-parameters}
We adopt the Python MARL framework (PyMARL) \cite{samvelyan_starcraft_2019} on the github to develop our algorithm. The hyper-parameters of training and testing configurations are the same as in SMAC \cite{samvelyan_starcraft_2019} and could be referred in the source codes. Here we list the parameters of Qatten's mixing network in Table~\ref{tab:para-qatten}. 

\subsection{Win Percentage Table of All Maps}
We here give the median win rates of all methods on the scenarios presented in our paper. In MMM2 and 3s5z\_vs\_3s6z, we augment Qatten with the weighted head Q-values. Other maps are reported by the win rates based on the original Qatten (Qatten-base). We could see Qatten obtains the best performance on almost all the map scenarios.

\begin{table}[htbp]
\footnotesize
\centering
  \caption{Median performance of the test win percentage.}
  \label{tab:all-results}
  \begin{tabular}{ccccccc}
    \toprule
    \multicolumn{1}{c}{Scenario} & 
    \multicolumn{1}{c}{\begin{tabular}[c]{@{}c@{}} Qatten \end{tabular}} & 
    \multicolumn{1}{c}{\begin{tabular}[c]{@{}c@{}} QMIX \end{tabular}} & 
    \multicolumn{1}{c}{\begin{tabular}[c]{@{}c@{}} COMA \end{tabular}} & 
    \multicolumn{1}{c}{\begin{tabular}[c]{@{}c@{}} VDN \end{tabular}} & 
    \multicolumn{1}{c}{\begin{tabular}[c]{@{}c@{}} IDL \end{tabular}} &
    \multicolumn{1}{c}{\begin{tabular}[c]{@{}c@{}} QTRAN \end{tabular}} \\
    \midrule
    2s\_vs\_1sc & \textbf{100} & \textbf{100} & 97 & \textbf{100} & \textbf{100} & \textbf{100} \\
    2s3z & \textbf{97} & \textbf{97} & 34 & \textbf{97} & 75 & 83 \\
    3s5z & \textbf{94} & \textbf{94} & 0 & 84 & 9 & 13 \\
    1c3s5z & \textbf{97} & 94 & 23 & 84 & 11 & 67 \\ \hline
    5m\_vs\_6m & \textbf{74} & 63 & 0 & 63 & 49 & 57 \\
    3s\_vs\_5z & \textbf{96} & 85 & 0 & 87 & 43 & 0 \\
    bane\_vs\_bane & 97 & 62 & 40 & 90 & 97 & \textbf{100} \\
    2c\_vs\_64zg & \textbf{65} & 45 & 0 & 19 & 2 & 10 \\ \hline
    MMM2 & \textbf{79} & 61 & 0 & 0 & 0 & 0 \\
    3s5z\_vs\_3s6z & \textbf{16} & 1 & 0 & 0 & 0 & 0 \\
  \bottomrule
\end{tabular}
\end{table}

In the hard scenarios, 5m\_vs\_6m requires the precise control such as focusing fire and consistent walking to win. In 3s\_vs\_5z, since Zealots counter Stalkers, the only winning strategy is to kite the enemy around the map and kill them one after another. In 2c\_vs\_64zg, 64 enemy units make the action space of the agents the largest among all scenarios. In bane\_vs\_bane, the strategy of winning is to correctly use the Banelings to destroy as many enemies as possible. For super hard scenarios, the key winning strategy of MMM2 is that the Medivac heads to enemies first to absorbing fire and then retreats to heal the right ally with the least health. As the most difficult scenarios, 3s5z\_vs\_3s6z may require a better exploration mechanism while training.

\section{Extended Attention Analysis}
\label{sec:extended_analysis}
\begin{figure}[htbp]
\centering
\subfigure[\scriptsize{5m\_vs\_6m Attention Heat Map}]{
\label{3s5z-3s6z-head0}
\includegraphics[width=0.48\textwidth]{figures/attention/attention_heat_map_5m_vs_6m.pdf}}
\subfigure[\scriptsize{5m\_vs\_6m Agent Property}]{
\label{3s5z-3s6z-head1}
\includegraphics[width=0.48\textwidth]{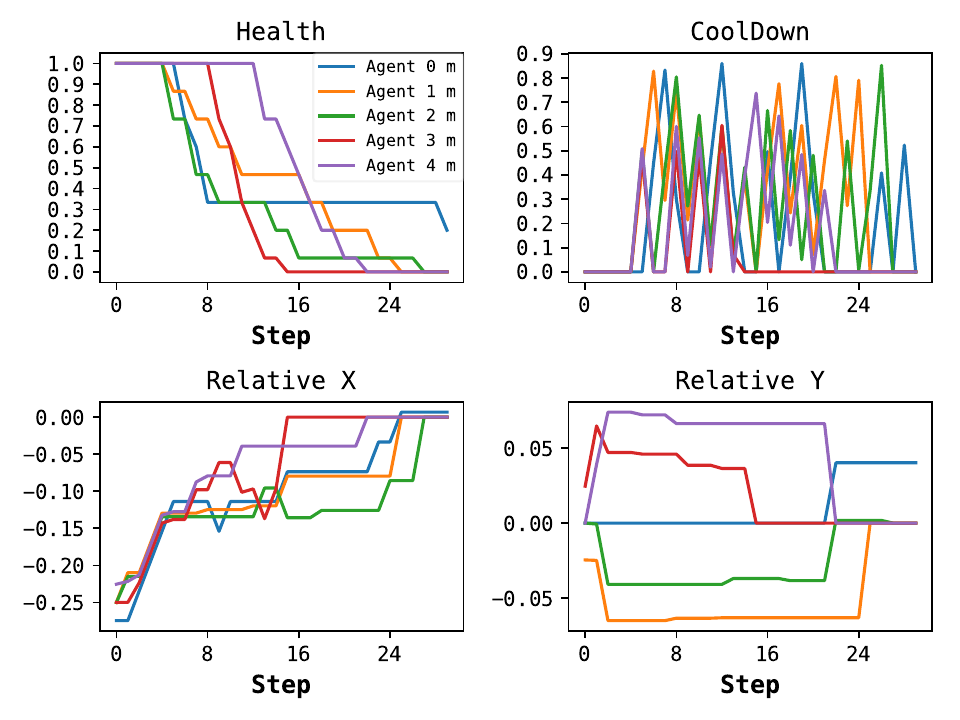}}
\subfigure[\scriptsize{3s5z\_vs\_3s6z Attention Heat Map}]{
\label{3s5z-3s6z-head2}
\includegraphics[width=0.48\textwidth]{figures/attention/attention_heat_map_3s5z_vs_3s6z.pdf}}
\subfigure[\scriptsize{3s5z\_vs\_3s6z Agent Property}]{
\label{3s5z-3s6z-head3}
\includegraphics[width=0.48\textwidth]{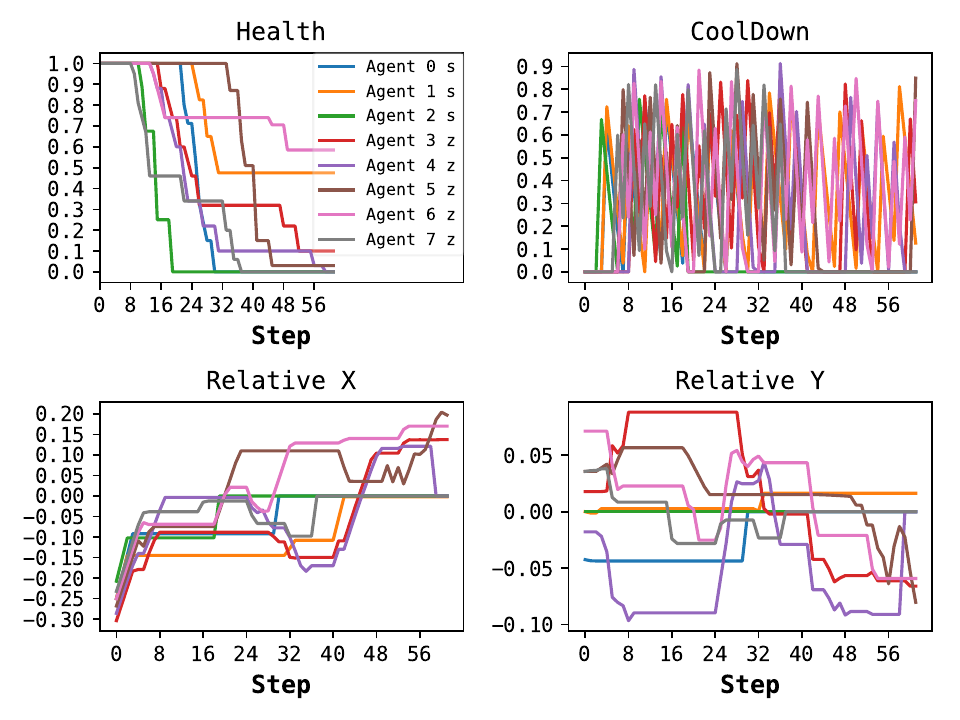}}
\caption{Attention weights on 5m\_vs\_6m and 3s5z\_vs\_3s6z. Steps increase from top to the bottom in the attention heat maps. Horizontal ordination indicates the agent id under each head.} 
\label{figure:attention_extended}
\end{figure}

For 5m\_vs\_6m, the most important trick to win is to avoid being killed and focusing fire to kill enemy. From Figure~\ref{figure:attention_extended}(a), we see allies have similar attention weights, which means that Qatten indicates an almost equally divided $Q_{tot}$ for agents. This is because each ally marine plays similar roles on this homogeneous scenario. This may also explain the fact that VDN could slightly outperform QMIX when the weights of each $Q^{i}$ is roughly equal. But we still find some tendencies of Qatten. One main tendency is that, the only alive unit (agent 0) receives the highest weight in the later steps of the episode, which encourages agents to survive for winning. With the refined attention mechanism adjusting weights of $Q^{i}$ with minor difference, Qatten performs the best.

We analyse another map 3s5z\_vs\_3s6z and the attention heat map is shown in Figure~\ref{figure:attention_extended}(c). After analyzing the correlation of attention weights and agent features, we notice that Head 1 focuses on the output damage. Units with high CoolDown (indicating just fired) receives high attention values. Head 1 and 2 tend to focus on the Stalker agents (Agent 0, 1 and 2) while Stalkers (Agent 3-7) almost receives no attention at the beginning. While Stalker agent 1 with high health receives attention from head 1 and 2, Zealot agent 6 with high health receives attention from head 0 and 3. In summary, attention weights on map 3s5z\_vs\_3s6z are changing violently. As the battle involves highly fluctuant dynamics, VDN which statically decomposes $Q_{tot}$ equally thus cannot adapt to the complex scenarios. Similarly, QMIX with a rough and implicit approximation way also fails. In contrast, Qatten could approximate the sophisticated relations between $Q^{i}$ and $Q_{tot}$ with different attention heads to capture different features in sub-spaces.

\end{document}